\documentclass[final,twocolumn,authoryear,12pt]{elsarticle}

\usepackage{epsfig}

\usepackage{amssymb}

\usepackage[ps2pdf,%
a4paper=true,%
breaklinks=true,%
colorlinks=true,%
pdfauthor={First Author et al.},%
pdftitle={Simulation of the HiSCORE Experiment}%
]{hyperref}

\journal{Advances in Space Research}

\begin{document}

\begin{frontmatter}



\title{The ground-based {large-area} wide-angle $\gamma$-ray and cosmic-ray experiment \mbox{HiSCORE}}


\author{Martin Tluczykont\corref{cor}}
\ead{martin.tluczykont@physik.uni-hamburg.de}
\author{Daniel Hampf, Dieter Horns, Tanja Kneiske, Robert Eichler, Rayk Nachtigall}
\address{Department of Physics, University of Hamburg, Luruper Chaussee 149, 22761 Hamburg}
\cortext[cor]{Corresponding author}


\author{Gavin Rowell}
\address{University of Adelaide 5005, School of Chemistry \& Physics, Australia}

\begin{abstract}

{The question of the origin of cosmic rays
and other questions} of astroparticle and particle physics can be addressed with
indirect air-shower observations above 10\,TeV primary energy.
   We propose to explore the cosmic ray and $\gamma$-ray sky (accelerator sky) in the energy range from
   10\,TeV to 1\,EeV
   with the new ground-based large-area wide angle ($\Delta\Omega\,\sim$0.85\,sterad)
   air-shower detector \mbox{HiSCORE} (Hundred*i Square-km Cosmic ORigin Explorer).
   The \mbox{HiSCORE} detector is based on non-imaging air-shower Cherenkov light-front sampling using
   an array of light-collecting stations. A full detector simulation and basic reconstruction
   algorithms {have been used to assess the performance of HiSCORE.}
   First prototype studies for different hardware components of the detector array {have been carried out}.
   The resulting sensitivity of \mbox{HiSCORE} to $\gamma$-rays
   {will be comparable to CTA at 50\,TeV and will extend the sensitive energy range for $\gamma$-rays
   up to the PeV regime.
   \mbox{HiSCORE} will also be sensitive to charged cosmic rays between 100\,TeV and 1\,EeV.}
\end{abstract}


\end{frontmatter}

\parindent=0.5 cm

\section{Introduction}
The fundamental question of the origin of charged cosmic rays {at knee energies} remains unsolved.
Indirect air-shower observations of ultra-high energy $\gamma$-rays (UHE $\gamma$-rays, E\,$>$\,10\,TeV)
and cosmic rays above 100\,TeV are the key
to the solution of this question.
Additionally,
fundamental questions of particle physics
can be addressed with the same air-shower data and might partly have an
influence on its interpretation.
Among these particle-physics topics are
the measurement of the proton-proton cross-section,
search for quark-gluon plasma in air-showers,
axion search in the Galactic magnetic field,
search for Lorentz invariance violation
and for heavy super-symmetric particles (wimpzillas).

\paragraph{Cosmic rays}
{HiSCORE will provide spectral and composition measurements of cosmic rays over four decades in energy,
from 100\,TeV to 1\,EeV.}
In the past decades, many experiments {have measured} the energy spectrum and chemical composition of cosmic rays
over a wide range in energy.
Existing data suggests a Galactic origin of cosmic rays up to $\approx$\,10$^{17}$\,eV
\citep[e.g.][for reviews]{2003APh....19..193H,2009PrPNP..63..293B}.
The transition from a Galactic to an extragalactic origin of the observed
cosmic rays is believed to occur in the energy range between
10$^{15}$\,eV to 10$^{17}$\,eV.
Above 10$^{17}$\,eV{,}
composition is very poorly understood and compositions ranging from proton to iron dominated primaries are reported
\citep[][and references]{2004AnPhy.314..145A}.
A cosmic ray anisotropy on a large scale was observed in the northern hemisphere at TeV energies
by the Tibet air-shower array \citep{2006Sci...314..439A} and confirmed by the Super-Kamiokande-I detector
\citep{2007PhRvD..75f2003G} and Milagro \citep{2009ApJ...698.2121A}. Observations with IceCube yield
a consistent structure in the southern sky \citep{2010ApJ...718L.194A}.
The origin of this anisotropy remains uncertain.

\paragraph{Origin of Galactic cosmic rays}
The presence of CR accelerators {throughout} our Galaxy was clearly demonstrated by
observations of diffuse {$\gamma$-ray} emission along the Galactic plane by EGRET \citep{1997ApJ...481..205H}
and by later detections of extended $\gamma$-ray emission at VHE
\citep{2006Natur.439..695A,2007ApJ...658L..33A,2008A&A...481..401A}.
These emissions
are most likely produced in interactions of energetic CRs with
{diffuse gas.}
{Such extended molecular gas cloud structures thus act as CR tracers} \citep[see also][]{2007ApJ...665L.131G}.
The observation of UHE $\gamma$-ray emission from extended gas clouds ``illuminated'' by
nearby cosmic ray sources might open up a possibility to map the Galactic
cosmic ray {energy density.}
{Assuming} that the origin of cosmic rays is Galactic up to 10$^{17}$\,eV,
there must be objects within our Galaxy that accelerate cosmic rays up to PeV energies: the cosmic ray pevatrons.
Due to the typical inelasticity {of neutral pion production in proton--proton collisions}, pevatrons are expected to exhibit $\gamma$-ray spectra up to
several 100\,TeV.
Shell-type supernova remnants (SNR) are believed to be the prime candidates for the acceleration of Galactic cosmic rays.
However, {in the TeV energy regime}{,} the observed emission is ambiguous
since it can also be explained in a leptonic (inverse Compton) radiation scenario
{\citep[see][for a review]{2009MNRAS.392..240M}.}
{Recent results on the shell-type SNR RX\,J1713-3946 from the Fermi satellite
\citep{2011ApJ...734...28A} yield a hard spectrum in the GeV energy regime.
While these observations support a leptonic emission mechanism,
they might also be explained as hadronic emission
from dense cloud clumps \citep[][]{2010ApJ...708..965Z},
or when taking into account cosmic ray diffusion into mocelcular clouds \citep{2007ApJ...665L.131G}.
Also see \citep{2011arXiv1106.3080I}, for a discussion of thermal X-rays from RX\,J1713-3946.
}
Furthermore, the $\gamma$-ray luminosities of the detected objects are lower than expected,
and the observed spectra are soft or have cut-off energies in the TeV energy regime,
i.e. these objects cannot {currently} be cosmic ray pevatrons.
Finding the cosmic ray pevatrons requires a survey of a large part of the sky in the UHE $\gamma$-ray regime.
Interestingly, the leptonic/hadronic ambiguity disappears in the UHE regime,
where the inverse Compton scattering cross-section
{drops with increasing center of mass energy (Klein-Nishina regime).}
At 100\,TeV, the inverse Compton effect takes place in the deep Klein-Nishina regime
for electrons scattering off photons from the cosmic microwave background (CMB).
This results in inevitably soft $\gamma$-ray spectra from leptonic accelerators beyond 10\,TeV.
As opposed to that, a hard $\gamma$-ray spectrum continuing up to few hundred TeV
would be a clear signature of hadronic acceleration.

\paragraph{Origin of extragalactic cosmic rays}
Assuming the origin of cosmic rays beyond 10$^{17}$\,eV is extragalactic,
an enhancement of cosmic rays {beyond this energy} from the direction of the local super-cluster
can be expected \citep{2010cosp...38.2720K}.
These cosmic rays can interact with the
CMB, initiating intergalactic {secondary} cascades that could be observable in $\gamma$-rays.
The emission is expected to match the {structure of the local} super-cluster
and would be measurable as anisotropic emission in the total field of view of the experiment.
Alternatively, the accelerators of {extragalactic} cosmic rays might exhibit point-like
emission or halo-like structures resulting from the interactions of the accelerated particles
with the surroundings of the source.

\paragraph{Gamma-ray propagation and the hidden sector}
At such high energies as considered here,
$\gamma$-rays are attenuated via e$^+$e$^-$-pair-production with the photons of low energy radiation fields,
such as the cosmic microwave background (CMB), the extragalactic background light, the supergalactic radiation field or
the Galactic interstellar radiation fields (IRF).
Within our Galaxy, the dominant radiation fields are the IRF and the CMB.
The attenuation
reaches a maximum for Galactic objects around 100\,TeV from the Galactic IRF and at 2\,PeV from the CMB
\citep{2006ApJ...640L.155M}.
While the former depends on the Galactic longitude and local radiation fields, the latter is universal.
If the distance of the observed $\gamma$-ray sources is known, the density of the IRF might
be inferred from the strength of the absorption by pair production (or from spectral features).
Inversely, Galactic absorption might also open up a new possibility
to infer distances from the measurement of $\gamma$-ray spectra, if
the IRF in the line of sight is known. Such a new method for distance estimation of Galactic
objects might also be possible if the universal absorption by the CMB could be measured.
The expected attenuation by pair production might be altered by photon/axion
conversion \citep[e.g.][]{2009EPJC...59..557S}. Photons produced
at the source travelling through the Galactic magnetic field might convert into axions propagating
without absorption. If a reconversion of these axions back into photons happens before arriving at Earth,
the photon signal would appear to be stronger than expected.
The same effect could arise if photon/hidden-photon oscillations \citep{2008AIPC.1085..727Z} would occur.
Another effect that might alter the expected absorption by pair production is
the modification of the e$^+$e$^-$ pair production threshold in case of
Lorentz invariance violation.

\subsection{A new non-imaging {UHE} detector: \mbox{HiSCORE}}
The sensitivity level of existing and currently planned $\gamma$-ray detectors
is optimized to the very high-energy regime (VHE, 100\,GeV\,$<$\,E\,$<$\,10\,TeV).
The sensitivity to the UHE $\gamma$-ray regime (UHE $\gamma$-rays, E\,$>$\,10\,TeV)
is limited because previously, the trend in development of detectors for
$\gamma$-ray astronomy was dominated by the focus on low energy thresholds.
Future and planned experiments such as CTA
\citep{2010arXiv1008.3703C}, HAWC \citep{2005AIPC..745..234S}, TenTen \citep{2008AIPC.1085..813R},
or LHAASO \citep{2010arXiv1006.4298C} will improve
the situation in the UHE $\gamma$-ray regime.
However, due to dropping event statistics with rising energy,
the key to UHE $\gamma$-ray astronomy is a very large instrumented
area of the order of 10 to 100\,km$^2$.
While such large instrumented areas seem impracticable using the well-established
imaging air Cherenkov technique (order of 10\,000 channels/km$^2$),
the non-imaging air Cherenkov technique provides a complementary possibility
that comes along with some advantages, such as a small number of channels per km$^2$
(less than 100 per km$^2$) and a wide field of view (order of sr).
We have started the development of \mbox{HiSCORE} (Hundred Square-km Cosmic ORigin Explorer),
a ground-based wide-angle large-area air-shower detector
for non-imaging $\gamma$-ray astronomy and cosmic ray physics from 10\,TeV to 1\,EeV.
With its wide field of view (continuously monitoring a large part of the sky) and
a focus on the highest energies, this project is complementary to
(yet independent of) existing and planned experiments.

Optimized for the UHE $\gamma$-ray regime and for cosmic ray energies from 100\,TeV
to 1\,EeV, \mbox{HiSCORE} will allow to address the $\gamma$-ray and cosmic ray physics topics introduced
in the previous section.
%

\section{\mbox{HiSCORE} detector design}
The \mbox{HiSCORE} detector principle is based on the shower front sampling technique
using Cherenkov light.
The detector consists of a large array of wide-angle light-sensitive detector
stations, that measure the light amplitude and the entire longitudinal development
using the shower-front arrival-time distribution
(at distances from the shower core $>$100\,m).
These measurements allow detailed spectral and composition measurements,
and $\gamma$-hadron separation via reconstruction of the
shower-depth.
The concept of the detector modules used as working
assumption for the \mbox{HiSCORE} detector was adapted from previous $\gamma$-ray
experiments, such as
THEMISTOCLE \citep{1990NuPhS..14...79F,1991AIPC..220..237B},
HEGRA AIROBICC \citep{1995APh.....3..321K}, or BLANCA \citep{1997ICRC....5..189C}.
Similar detector modules are also used in the
TUNKA array for cosmic-ray physics \citep{2005ICRC....8..255B}.
As compared to TUNKA, three aspects of \mbox{HiSCORE} will
be different: an instrumented area larger by more than an order of magnitude,
larger detector station areas (factor 16) and larger inter-station spacing.
Advances made in technology allow improvements
to the original detector components such as
improved photo-sensitive detectors,
and fast trigger and readout electronics, therewith allowing
a measurement of the Cherenkov photon arrival time distribution.

\begin{figure}[!t]
  \begin{center}
  \includegraphics[width=\columnwidth]{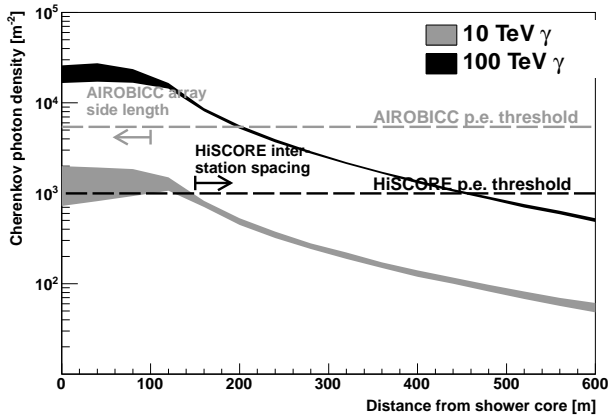}
  \end{center}
  \caption{Lateral photon density function (LDF) of Cherenkov light {(300\,nm to 600\,nm)} {at} sea level for airshowers
           initiated by a $\gamma$-ray at 10\,TeV and 100\,TeV.
           The sensitivity level of one \mbox{HiSCORE} detector station is indicated by the {dashed} black line.
           For comparison, the corresponding sensitivity level for AIROBICC is also shown.}
  \label{LDF}
 \end{figure}

A very large instrumented area is required
to reach sufficiently large event statistics {in the UHE regime},
and is achieved with a low array density, i.e. large inter-station spacings.
A reasonable value for the detector station spacing can be
derived from the lateral photon density function (LDF) of Cherenkov light {(300\,nm to 600\,nm)} at
observation level, shown in Figure~\ref{LDF}.
Within a radius of 120\,m around the shower core position
the LDF is roughly constant, but shows
a large spread from shower to shower. Fluctuations are much lower
beyond 120\,m.
With the envisaged station spacing of 100--200\,m, only few stations
{are} within the inner 120\,m of the LDF, shown in Figure~\ref{LDF}.
Thus, \mbox{HiSCORE} will primarily sample the outer part of the LDF.
The low photon density far away from the shower core justifies
the chosen large individual detector station areas.
For comparison, the corresponding sensitive range of the AIROBICC experiment is also shown
(grey {dashed} line).
This figure demonstrates the basic difference in scale: The inter-station spacing of
the \mbox{HiSCORE} array is of the same order of magnitude as the total side-length of AIROBICC.
Measurements of the LDF far away from the shower core provide a large lever-arm
and thus good reconstruction quality.
Another important aspect for the reconstruction with the \mbox{HiSCORE} detector
is the usage of the full timing information
from the arrival time distribution of the Cherenkov photons at each detector station.
The event reconstruction of \mbox{HiSCORE} is based on the combination of information from
the lateral photon density distribution and the arrival time distribution of Cherenkov photons
\citep{2009arXiv0909.0663H}.

\begin{figure}[ht!]
  \begin{center}
  \includegraphics[width=\columnwidth]{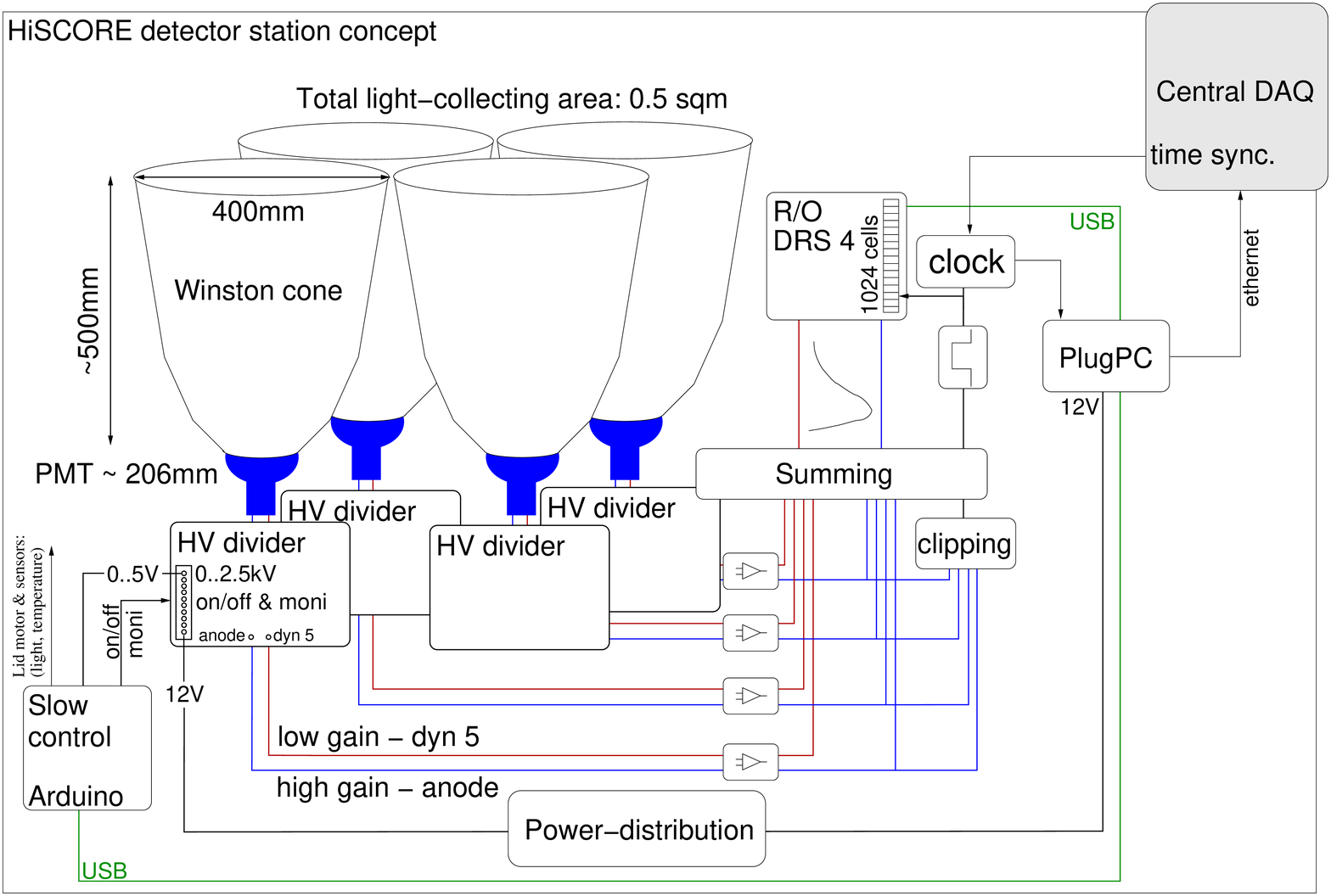}
  \end{center}
  \caption{\mbox{HiSCORE} detector station concept. The four PMTs with Winston cones and all electronics parts will be mounted inside
	   a station box equipped with a sliding lid.}
  \label{StationConcept}
 \end{figure}

Each detector station consists of four photomultiplier tubes (PMTs) equip\-ped with four
light-collecting Winston cones of 30$^\circ$ half-opening angle pointing to the zenith.
A schematical drawing of the station concept is shown in Figure~\ref{StationConcept}.
{Each PMT channel is equipped with an HV board (voltage supply and divider).
In addition to the anode signal (high gain) of the 6-stage PMTs, the signal at the 5th dynode is read out as well (low gain).}
All four modules (PMT+cone), including the trigger, readout electronics and communication {(also see next section)}
are planned to be encased in a box equipped with a sliding lid.
The advantages of using four PMT channels per station are the possibility
to suppress false triggers from nightsky background (NSB) light by a local coincidence trigger
condition and the resulting large light collecting area $a$.
A total area of $a$~=~0.5\,m$^2$ can be achieved when using four 8'' PMTs and a Winston cone height
of 0.5\,m.
A fast signal readout and digitization in the GHz regime are needed.
Different solutions
such as analog ring samplers or domino ring samplers (DRS) are under study.
We are currently testing the DRS\,4 chip that was developed by the PSI\footnote{{\tt http://midas.psi.ch/drs}}.

\section{\mbox{HiSCORE} simulation results}
\subsection{Air-shower and detector simulation}
Air showers were simulated with CORSIKAv675 \citep{1998cmcc.book.....H}
using the hadronic interaction model GHEISHA \citep{fesefeldt:1985a}.
{Showers initiated by primary} $\gamma$-rays, protons, Helium- Nitrogen- and Iron-nuclei were simulated
in the energy-range from 10\,TeV to 10\,PeV following
a powerlaw distribution with a spectral index of \mbox{-1}.
{Additionally, protons were simulated down to 5\,TeV (see discussion of hadron trigger rate below).}
The IACT option was used, storing Cherenkov photons in spheres of 1\,m radius at sea-level,
each sphere representing one detector.
The array layout was simulated as a simple grid of 22\,$\times$\,22 stations
with an inter-station spacing of 150\,m, covering a total instrumented area of $\approx$\,10\,km$^2$ (3.15\,km side-length).
Air-showers were simulated over a larger area, with random impact position in a square with side-length 3.75\,km.
A full detector simulation (\emph{sim\_score}) was implemented
on the basis of the \emph{iact} package provided by \citet{2008APh....30..149B}.
At the position of each CORSIKA sphere, a detector station with 4 PMT-channels
is simulated in \emph{sim\_score}. The detector station simulation includes
atmospheric absorption \citep[MODTRAN][]{modtran},
ray-tracing tables for Winston cone acceptance, and PMT response {(including afterpulses)}.
The simulation of the station trigger {(as illustrated in Figure~\ref{StationConcept})} includes clipping of each individual PMT signals {(to suppress the effect of afterpulses)}, analog summing of all four channels, and a discriminator.
A local station trigger is issued when the sum of all four PMT signals passes a given
threshold (few $\sigma$ above NSB level).
The resulting simulated digitized signals
are stored and a bin-by-bin noise pedestal from a simulation of the expected NSB
is added.

\subsection{Cosmic ray trigger rates}
\begin{figure}
\begin{center}
\includegraphics[width=\columnwidth]{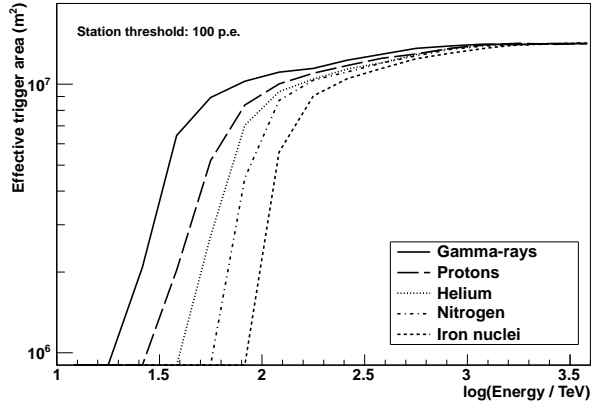}
\end{center}
\caption{Effective areas $A_{eff}$ of the \mbox{HiSCORE} detector at trigger level
         for different primary particle types. {An array trigger is issued when one or more
	 stations have a local trigger.}}
\label{aeff_trigger}
\end{figure}

The effective area at trigger level is given as the ratio of triggered to simulated events divided by
the simulated area.
Figure~\ref{aeff_trigger} shows the effective trigger areas (for $\gamma$s, H, He, N, Fe)
when using a station threshold of 100\,photoelectrons (p.e.).
{As can be seen from the effective trigger areas, protons still trigger at low energies as opposed to
heavier nuclei. This justifies the simulation of proton energies down to 5\,TeV.}
Using these effective trigger areas and the \mbox{\emph{polygonato}} parametrization
for cosmic rays \citep{2003APh....19..193H},
a cosmic ray trigger rate of $\approx$\,1.8\,kHz was obtained for a 10\,km$^2$ array.
This corresponds to a local single-station {cosmic ray} trigger rate of $\approx$\,15\,Hz.
The data rate could be further reduced when using a two-fold next-neighbour station coincidence
condition.
Such a coincidence condition may be implemented at the software-level of the central data acquisition or using
a hardware second-level trigger condition.

A simulation of the expected accidental {local station}
trigger rate due to night-sky background (NSB) photons {
(PMT response including afterpulses) was implemented.
Results from measurements of the NSB photon rate
in Australia \citep{2011arXiv1105.1251H} were used as input and a discriminator response gate with a width of 7\,ns was assumed.}
{At a discriminator threshold of 100\,p.e., this simulation} yielded an NSB trigger rate of the order of
{100\,Hz}.
This value clearly demonstrates that, at a station threshold of 100\,p.e.,
the station trigger rate is dominated by NSB photons.
{
First tests of the planned readout system yield a maximum data rate of the order of 100\,Hz per station.
Slightly increasing the station discriminator threshold to 105\,p.e. results in an NSB trigger rate of 52\,Hz
without significantly affecting the sensitivity at reconstruction level ($\ge$\,3\,Stations and reconstruction cuts).}

For the 10\,km$^2$ stage of the array,
the trigger rates (E$>$\,5TeV) of the different simulated cosmic ray particle-classes
are 875\,Hz (Z=1), 505\,Hz (Z=2-5), 290\,Hz (Z=6-24), and 103\,Hz (Z$>$24).
{In this detector stage,} we expect {of the order of 10$^{8}$} cosmic ray events per year above 100\,TeV
and still of the order of {5 events} per year at {$10^{18}$\,eV}.
A $\gamma$-ness parameter\footnote{The $\gamma$-ness parameter is defined
on the basis of the reconstructed shower depth and energy and the light concentration on the ground.}
\citep{2009arXiv0909.0663H} is used for
$\gamma$-hadron separation and can also be used for a measurement of the chemical composition, ultimately providing an
estimation of the mass-composition via the measurement of the shower depth. A similar approach is
used by the TUNKA detector \citep{2005ICRC....8..255B}.

\subsection{Gamma-ray sensitivity}
The point-source survey sensitivity of \mbox{HiSCORE} to $\gamma$-rays
was calculated using basic reconstruction algorithms introduced in
\citet{2009arXiv0909.0663H}.

The performance of the reconstruction algorithms 
as applied to the Monte Carlo data set is summarized in Table~\ref{performance}.
\begin{table*}[ht]
\caption{\label{performance}
	Performance of the {employed} basic reconstruction algorithms {for $\gamma$-ray analysis}.
	}
\centering
\begin{tabular}{llll}
\hline
					& 50\,TeV	& 100\,TeV		& $>$500\,TeV		\\
Angular resolution	(68\,\%)	& 0.8$^\circ$	& 0.3$^\circ$		& 0.1$^\circ$		\\
Shower core position accuracy		& 40\,m		& 20\,m			& 8\,m			\\
Shower depth accuracy			& 150\,g/cm$^2$	& 60\,g/cm$^2$		& 20\,g/cm$^2$		\\
Energy resolution	($\Delta E / E$)& 40\,\%	& 20\,\%		& 10\,\%		\\
\hline
\end{tabular}
\end{table*}
The fundamental differences between \mbox{HiSCORE} and an imaging air-shower array such as CTA are the much larger
detector area of \mbox{HiSCORE} and the fact that \mbox{HiSCORE}, with its large field of view,
will always simultaneously cover a large fraction of the sky.
Therefore, \mbox{HiSCORE} always operates in \emph{survey mode},
and any single source inside the large field of view will be visible over 200\,h per year
(calculated for a southern hemisphere site at a latitude of -35$^\circ$),
i.e. 1000\,h after 5\,years of survey operation.
As opposed to that, IACTs operate in pointed mode and cannot allocate such long observation times to single sources.
For example, in the H.E.S.S. survey \citep{2006ApJ...636..777A} the time allocated to one {pointing per year}
is typically of the order of 10\,h, i.e. {a factor of 20 lower than} in the \mbox{HiSCORE} survey.
{At the same time, with a covered solid angle of $\pi$\,steradian for more than 200\,h per year,
HiSCORE covers a significantly larger fraction of the sky than the H.E.S.S. survey has covered.}
%

\begin{figure}[ht]
\begin{center}
\includegraphics[width=\columnwidth]{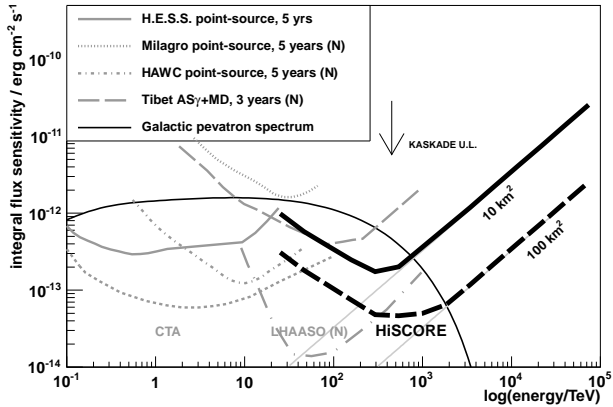}
\end{center}
\caption{
                \mbox{HiSCORE} 5-year point-source survey sensitivity for 10\,km$^2$ and 100\,km$^2$ instrumented detector areas.
                For comparison, {point-source sensitivities are given for the following experiments:
		H.E.S.S (50\,h), CTA (50\,h), Milagro and HAWC (5-years), Tibet AS$\gamma$+MD (3\,years), and LHAASO.
                In a survey-mode, the CTA and H.E.S.S. sensitivities will not reach the displayed level
		but will be significantly less powerful, because pointed instruments}
                cannot cover as large a fraction of the sky as \mbox{HiSCORE}.
		{Instruments located or planned to be located in the northern hemisphere are marked with (N).}
         }
\label{sensitivity}
\end{figure}

To calculate the sensitivity to $\gamma$-rays, we required 5\,$\sigma$ detection significance and a minimum of 50 $\gamma$-rays.
The background was calculated on the basis of the simulated hadron effective areas
after applying reconstruction cuts (at least 3 triggered stations, $\gamma$-ness\,$>$\,6, contained events).
The \emph{polygonato} parametrization \citep{2003APh....19..193H}
of the cosmic ray spectra for each nucleus-type was folded with the area corresponding to the closest nucleus
out of the group of four simulated (proton, Helium, Nitrogen, Iron).
The resulting point-source sensitivities for instrumented detector areas of 10 and 100\,km$^2$ are shown in Figure~\ref{sensitivity}
along with point-source sensitivities {of other experiments \citep{2008AIPC.1085..874B,2010arXiv1008.3703C,2007Ap&SS.309..435A,hawchome},}
 and an upper limit by KASKADE \citep{2004ApJ...608..865A}.
{HiSCORE will be complementary to other instruments in different ways. It will extend the energy range covered by CTA at
a similar energy flux sensitivity level and cover a large fraction ($\pi$\,sr) of the southern sky.
While the 10--100\,km$^2$ stages of HiSCORE
are planned for deployement in the southern hemisphere (best access to Galactic plane), the other survey instruments shown
here (AS$\gamma$+MD and LHAASO) are northern hemisphere detectors, thus covering a different region of the sky.}

At the energy threshold, the \mbox{HiSCORE} sensitivity is limited by the angular resolution and the $\gamma$-hadron separation.
At the upper energy end, the sensitivity is statistics limited and only depends on the total detector area and the exposure time
(grey thin rising straight lines).
{Weak pevatrons (thin black line) {that might be detectable below their cut-off energy regime} by H.E.S.S. and Milagro
will be {well} within reach of a 10\,km$^2$ detector. The cut-off regime of such sources will be fully
resolved with \mbox{HiSCORE} (100\,km$^2$).}

\section{Prototype developement and engineering array}
A prototype station is currently under developement and hardware component tests are in progress.
A test-bed for photomultiplier-tubes (PMTs) was developed.
We are currently measuring signal shapes, gain values, and dynamic ranges.
The dynamic range will be increased to the necessary factor of 10$^5$ (10\,TeV to 1\,EeV)
by reading out one or two dynodes in addition to the anode signal.
A first prototype station with a single PMT-channel is currently under construction at the University of Hamburg.
The aluminium housing is equipped with a sliding lid and also contains slow-control electronics,
a high-voltage supply and distribution, and a read-out system.
We plan the deployment of full (4-channel) prototype stations for field tests {and cross-calibration} on
the Tunka-site in Russia. Observatory sites in the southern hemisphere,
such as the AUGER site, are also considered
for test deployements. Especially in view of the construction of a first stage of the \mbox{HiSCORE} array,
a site in the southern hempisphere is interesting due to the better visibility of the Galactic plane.
{Relative timing accuracy can be a limiting factor for the quality of the reconstruction. We aim at
a 1\,ns relative time resolution.
An already existing time-calibration will be available on the Tunka site, based on the usage of the carrier
frequency of optical fibers (also used for readout). An improvement of the Tunka method to an accuracy
of 1\,ns is planned. Furthermore, an investigation of alternative time-calibration methods, such as radio-beacon
usage, are planned \citep{2010NIMPA.615..277S}.}

\section{Summary and outlook}
We propose to explore the accelerator sky with observations of cosmic rays (100\,TeV to 1\,EeV)
and UHE $\gamma$-rays (E$>$10\,TeV) with the new wide-angle large-area
air-shower detector \mbox{HiSCORE}.
Fundamental questions of particle and astroparticle physics can be addressed with \mbox{HiSCORE}.

HiSCORE will open up the {UHE} $\gamma$-ray (E$>$10\,TeV) observation window.
{Around 50\,TeV a sensitivity comparable to the currently planned CTA will be reached.
HiSCORE will extend the sensitive energy range up to the PeV regime.}
Measurements of the cosmic ray spectrum and mass composition will be possible
over 4 orders of magnitude, from 100\,TeV to 1\,EeV.

The final layout of the \mbox{HiSCORE} detector is under study. 
Varying detector station densities can be used to optimize the sensitivity over the whole energy range.
This could be achieved using smaller cells with very small inter-station distances (order of 10\,m)
and a graded array structure, with increasing inter-station distances towards the array-edge.
An enhancement of the duty-cycle (10\,\%, air-Cherenkov observations restricted to astronomical darkness)
could be achieved by equipping the underside of the sliding station lids with scintillator material,
thus providing charged particle air-shower measurements during daytime.
Furthermore, a possible combination of the \mbox{HiSCORE} station principle with imaging air Cherenkov telescopes
is under study.

\clearpage
\end{document}